\documentclass[prb,aps,twocolumn,superscriptaddress,floatfix]{revtex4-2}

\usepackage{amsmath}
\usepackage{amssymb}
\usepackage{natbib}
\usepackage{hyperref}
\usepackage{graphicx}
\usepackage{dcolumn}
\usepackage{bm}
\usepackage[mathlines]{lineno}
\usepackage[dvipsnames]{xcolor}
\usepackage{tikz}
\usepackage{tikz-3dplot}
\usepackage{ifthen}
\usepackage{multirow}
\usepackage{pgfplots}
\pgfplotsset{compat=1.16}
\usepackage{comment}
\usepackage{soul}
\usepackage[normalem]{ulem}

\newcommand{\estado}[1] {\left| #1 \right\rangle}

\begin{document}
	
	\title{Magnetic structure and component-separated transitions of HoNiSi$_{3}$}
	
	\author{R. Tartaglia}
	\affiliation{Universidade Estatual de Campinas (UNICAMP), Instituto de F\'isica Gleb Wataghin, Campinas, S\~ao Paulo 13083-859, Brazil}
	
	\author{F. R. Arantes}
	\affiliation{CCNH, Universidade Federal do ABC (UFABC), 09210-580, Santo Andr\'e, S\~ao Paulo, Brazil}
	
	\author{C. W. Galdino}
	\altaffiliation[Present address: ] {Swiss Light Source, Paul Scherrer Institute, 5232 Villigen-PSI, Switzerland}
	\affiliation{Universidade Estatual de Campinas (UNICAMP), Instituto de F\'isica Gleb Wataghin, Campinas, S\~ao Paulo 13083-859, Brazil}

	\author{U. F. Kaneko}
	\affiliation{Brazilian Synchrotron Light Laboratory (LNLS), Brazilian Center for Research in Energy and Materials (CNPEM), Campinas, S\~ao Paulo 13083-970, Brazil}
	
	\author{M. A. Avila} 
	\affiliation{CCNH, Universidade Federal do ABC (UFABC), 09210-580, Santo Andr\'e, S\~ao Paulo, Brazil}
	
	\author{E. Granado} 
	\affiliation{Universidade Estatual de Campinas (UNICAMP), Instituto de F\'isica Gleb Wataghin, Campinas, S\~ao Paulo 13083-859, Brazil}
	
	\author{Veronica Vildosola}
	\affiliation{Consejo Nacional de Investigaciones Cient\'ificas y T\'ecnicas (CONICET), 1040 Ciudad Autonoma de Buenos Aires, Argentina}    
	\affiliation{Departamento de Materia Condensada, GIyA, CNEA San Mart\'in, 1650 Provincia de Buenos Aires, Argentina}
	\affiliation{Instituto de Nanociencia y Nanotecnolog\'ia CNEA-CONICET, 8400 Bariloche, Argentina}
	
	\author{Matias Nu\~nez}
	\affiliation{Consejo Nacional de Investigaciones Cient\'ificas y T\'ecnicas (CONICET), 1040 Ciudad Autonoma de Buenos Aires, Argentina}
	\affiliation{Instituto de Investigaciones en Biodiversidad y Medioambiente (INIBIOMA), Universidad Nacional del Comahue, 1415 Bariloche, Argentina}
	
	\author{Pablo S. Cornaglia}
	\affiliation{Consejo Nacional de Investigaciones Cient\'ificas y T\'ecnicas (CONICET), 1040 Ciudad Autonoma de Buenos Aires, Argentina}
	\affiliation{Instituto de Nanociencia y Nanotecnolog\'ia CNEA-CONICET, 8400 Bariloche, Argentina}
	\affiliation{Centro At\'omico Bariloche and Instituto Balseiro, CNEA, 8400 Bariloche, Argentina}
	
	\author{Daniel J. Garc\'ia}
	\affiliation{Consejo Nacional de Investigaciones Cient\'ificas y T\'ecnicas (CONICET), 1040 Ciudad Autonoma de Buenos Aires, Argentina}
	\affiliation{Centro At\'omico Bariloche and Instituto Balseiro, CNEA, 8400 Bariloche, Argentina}
	
	\begin{abstract}
		
		HoNiSi$_{3}$ is an intermetallic compound characterized by two successive antiferromagnetic transitions at $T_{N1} = 6.3$ K and $T_{N2} = 10.4$ K. Here, its zero-field microscopic magnetic structure is inferred from resonant x-ray magnetic diffraction experiments on a single crystalline sample that complement previous bulk magnetic susceptibility data. For $T < T_{N2}$, the primitive magnetic unit cell matches the chemical cell. The magnetic structure features ferromagnetic {\it ac} planes stacked in an antiferromagnetic $\uparrow \downarrow \uparrow \downarrow$  pattern. For $T_{N1} < T < T_{N2}$, the ordered magnetic moment points along $\vec{a}$, and for $T < T_{N1}$ a component along $\vec{c}$ also orders. A symmetry analysis indicates that the magnetic structure for $T<T_{N1}$ is not compatible with the presumed orthorhombic $Cmmm$ space group of the chemical structure, and therefore a slight lattice distortion is implied. Mean-field calculations using a simplified magnetic Hamiltonian, including a reduced set of three independent exchange coupling parameters determined by density functional theory calculations and two crystal electric field terms taken as free-fitting parameters, are able to reproduce the main experimental observations. An alternative approach using a more complete model including seven exchange coupling and nine crystal electric field terms is also explored, where the search of the ground state magnetic structure compatible with the available anisotropic magnetic susceptibility and magnetization data is carried out with the help of an unsupervised machine learning algorithm. The possible magnetic configurations are grouped into five clusters, and the cluster that yields the best comparison with the experimental macroscopic data contains the parameters previously found with the simplified model and also predicts the correct ground-state magnetic structure. 
		
	\end{abstract}
	
	\maketitle	
	
	\section{Introduction}
	
	The heavy rare-earth (\textit{R}) elements have rich magnetic phase diagrams with multiple phase transitions. For instance, Dy and Ho display helical antiferromagnetic (AFM) structures with propagation vectors along the hexagonal axis below $T_{N}= 179$ and $132$ K, respectively \cite{Behrendt1958,Koehler1966}. Upon further cooling, Dy orders ferromagnetically below $T_C= 85$ K whereas Ho develops a conic spiral structure below $T_C=20$ K. Such intriguing behavior results from a strong interplay between the long-range Ruderman-Kittel-Kasuya-Yosida (RKKY) exchange coupling, temperature-dependent crystal electric field (CEF) parameters, and also anisotropic magnetic dipole interactions in some cases \cite{jensen1991rare}. 
	
	As could be anticipated, some \textit{R}-based compounds also show intriguing properties, such as different components of the total magnetic moment ordering independently at different temperatures. This phenomenon has been observed in a few compounds such as DyB$_{4}$~\cite{Watanuki2005} and HoRh$_{2}$Si$_{2}$ \cite{Shigeoka2011}. DyB$_{4}$ crystallizes in a primitive tetragonal lattice, with space group \textit{P4/mbm}. At the Néel temperature $T_{N2} = 20.3$ K, a collinear AFM ordering with the magnetic moment oriented along the tetragonal $\vec{c}$ direction develops. Another AFM ordering occurs at $T_{N1} = 12.7$~K, where an \textit{ab} component of the magnetic moment orders \cite{Watanuki2005,Okuyama2005,Ji2007} accompanied by a slight monoclinic distortion \cite{Okuyama2005,Ji2007}. HoRh$_{2}$Si$_{2}$ has a body-centered tetragonal lattice (\textit{I4/mmm} space group). The higher-temperature phase transition at $T_{N2} = 29.5$~K is related to the AFM ordering of the Ho magnetic moments along the $\vec{c}$ axis. Below $T_{N1} = 11.0$ K, the ordered magnetic moments tilt away from the $\vec{c}$ axis, with the tilting angle being temperature-dependent and vanishing at $T_{N1}$ \cite{Shigeoka2011,Slaski1983,Sekizawa1987,Jaworska-Golab2002}. In these two systems, it is claimed that quadrupole interactions play a role in the occurrence of the split transitions \cite{Watanuki2005,Okuyama2005,Ji2007,Shigeoka2011,Shigeoka2015,Song2020}, since strong spin-orbit coupling correlates spin and orbital degrees of freedom, thus enabling the ordering of high order multipoles. On the other hand, a mean-field approximation with nearest-neighbor exchange interaction and CEF parameters up to fourth-order is sufficient to properly capture the macroscopic properties for both compounds at zero field  \cite{Matsumura2011,Takano1987,Jaworska-Golab2002}.
	
\begin{figure}[htb]
		\begin{center}
			\includegraphics[width=\columnwidth]{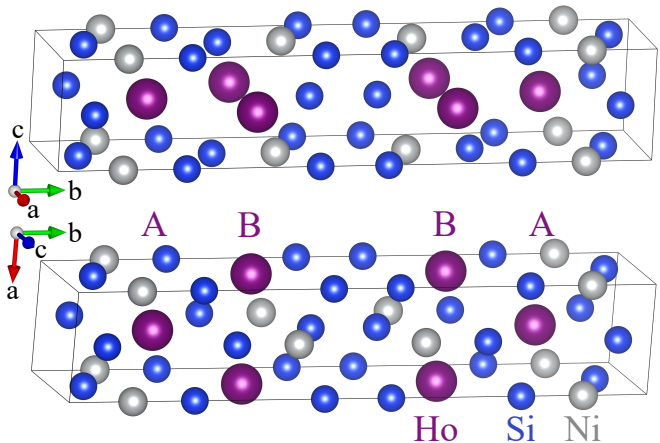}
			\caption{Crystal structure of HoNiSi$_3$ \cite{Arantes2018, konczyk2005erbium}, with solid black line defining the unit cell. The crystal structure of the \textit{R} ions can be seen as ABBA stacking of rectangular layers along the $\vec{b}$ direction.}
			\label{figcrystal}
		\end{center}
	\end{figure}

	The \textit{R}NiSi$_{3}$ (\textit{R} = Y, Gd-Lu) intermetallic series crystallizes in a $C$-centered orthorhombic lattice (\textit{Cmmm} space group, see Fig. \ref{figcrystal}). For $R=$ Gd and Tb, ferromagnetic (FM) {\it ac}~planes are found to be stacked antiferromagnetically in a $\uparrow \downarrow \uparrow \downarrow$ pattern below $T_{N} = 22.2$ and $33.2$~K, respectively, with spontaneous moments pointing along $\vec{a}$ \cite{Tartaglia2019}. Conversely, in YbNiSi$_{3}$ the stacking follows a $\uparrow \downarrow \downarrow \uparrow$ pattern with moments pointing along $\vec{b}$ \cite{kobayashi2008neutron}. The distinct stacking patterns of the magnetic end members of this series bring attention to the intermediary members $R =$ Dy-Tm, for which only macroscopic magnetic measurements are available so far \cite{Avila2004,Arantes2018}. The behavior of HoNiSi$_{3}$ is particularly interesting. Magnetic susceptibility and specific heat data reveal successive component-separated phase transitions at $T_{N1} = 6.3$~K and $T_{N2} = 10.4$ K associated with AFM ordering of the $\vec{c}$ and $\vec{a}$ moment components ($M_c$ and $M_a$, respectively). Whether or not higher multipole degrees of freedom are present and responsible for the features shown by HoNiSi$_{3}$, the natural subsequent step in the attempt to understand its ground state is determining its magnetic structure. Once the magnetic structure is resolved, additional constraints can be imposed on developing a theoretical microscopic model that describes the macroscopic data. 
	
	In this paper, we investigate the microscopic magnetism of HoNiSi$_{3}$ by combining a resonant x-ray magnetic diffraction experiment and magnetic simulations using both a simplified model and a complete set of exchange and CEF parameters. We find that the magnetic structures of both phases I ($T<T_{N1}$) and II ($T_{N1}<T<T_{N2}$) are commensurate with the chemical structure and share the same primitive unit cell, similar to GdNiSi$_{3}$ and TbNiSi$_{3}$. Also, representation analysis shows that the magnetic structure of phase II is described by a single irreducible representation of the \textit{Cmmm} space group. In contrast, two distinct irreducible representations are needed for phase I (one for each magnetic component), implying that a combined structural and magnetic phase transition must take place at $T_{N1}$. In fact, the magnetic space group symmetry is reduced from orthorhombic \textit{Cmmm'} in phase II to at least monoclinic \textit{C2'/m} in phase I. The detailed low-temperature magnetic orderings obtained in this paper and the thermodynamic measurements reported in Ref. \onlinecite{Arantes2018} allow us to compute possible exchange constants and CEF parameters.
	
	We also show how to combine the use of an unsupervised machine-learning algorithm to explore the whole parameter space of the magnetic Hamiltonian compatible with  the available thermodynamic measurements and find a set of possible ground-state magnetic structures. This approach could be particularly useful when the number of parameters is large and the experimental data are not sufficient to determine them uniquely.
	
	\section{\label{sec:Experimental} Resonant X-ray magnetic diffraction experiment}
	
	\subsection{\label{sec:Experiment} Experimental details}
	
	A platelet-shaped single crystal of HoNiSi$_{3}$ was grown from the melt in Sn flux as described elsewhere \cite{Arantes2018,Aristizabal-Giraldo2015}. Sample dimensions are $0.30\times 0.58 \times 0.15$ mm$^{3}$. Its largest natural face was employed in the measurements and corresponded to the crystallographic {\it ac} plane. Rocking curves of general $hkl$ reflections reveal mosaic widths between 0.02$^{\circ}$ and 0.04$^{\circ}$ full width at half maximum.
	
	Resonant x-ray diffraction measurements were performed at the x-ray diffraction and spectroscopy (XDS) beamline of the UVX ring of the Brazilian Synchrotron Light Laboratory in Campinas, with a 4 T superconducting multipolar wiggler source \cite{Lima2016}. The sample was mounted at the cold finger of a continuous-flow cryostat (base temperature 4.7 K) with a cylindrical Be window. The cryostat was attached vertically to the Eulerian cradle of a Huber 6+2 circle diffractometer appropriate for single-crystal x-ray diffraction, thus the probed scattering processes take place in the horizontal plane. The energy of the incident photons was selected by a double Si(111) crystal monochromator, with $L$N$_{2}$ cooling in the first crystal, whereas the second crystal was bent for sagittal focusing. The beam was vertically focused by a bent Rh-coated mirror placed downstream the monochromator, which also provided filtering of higher harmonics. The experiments were performed in the horizontal scattering plane, i.e., parallel to the linear polarization of the incident photons ($\pi$). A polarimeter stage was mounted upstream a scintillator detector, which enabled selecting either the $\pi \pi$' or $\pi \sigma$' polarization channels. For our experiments taken near the Ho $L_{3}$ edge, a Ge(333) analyzer was employed, yielding 2$\theta_{analyzer}$ = 89.66$^{\circ}$.
	
	\textit{R}-based magnetic compounds show strong dipolar resonances at the $L_{2,3}$ edges, reaching maximum intensities at energies $\sim 2$ eV above the corresponding edge positions \cite{Tartaglia2019,Granado2004,Lora-Serrano2006,granado2006magnetic}. As a preliminary x-ray fluorescence scan for HoNiSi$_3$ determined the Ho $L_{3}$ absorption edge to be at 8.074 keV (not shown), the photon energy was set at $8.076$ keV in our search for resonant magnetic reflections. In different runs, the sample was mounted in either \emph{AB} or \emph{BC} configurations, probing the $ab$ and $bc$ scattering planes, respectively [see the insets of Figs. \ref{Tdep}(a) and \ref{Tdep}(b)]. For dipolar resonances, the magnetic x-ray diffraction signal is sensible only to projections of the magnetic moment along the scattering vector \cite{Hill1996}. As previous magnetic susceptibility data indicate that there is no $\vec{b}$ component for the ordered Ho moment in HoNiSi$_{3}$ \cite{Arantes2018}, the $AB$ and $BC$ configurations probe the $\vec{a}$ and $\vec{c}$ components, respectively.
	
	\subsection{\label{sec:MagDiff}Results and analysis}
	
	A candidate magnetic structure of HoNiSi$_{3}$ would be the $\uparrow \downarrow \downarrow \uparrow$ stacking pattern along $\vec{b}$ such as found in YbNiSi$_3$ \cite{kobayashi2008neutron}, with propagation vector $\vec{k} = [1, 0, 0]$. In this case, the magnetic structure would break the $C$ centering of the charge crystal structure, and the magnetic reflections would be located in charge-forbidden $hkl$ positions of the reciprocal space with odd $h+k$. Attempts to observe such reflections at low temperatures $(T<T_{N1})$ in resonance condition were unsuccessful. In addition, 1D reciprocal space scans were performed along selected high-symmetry directions ([0,4,0] $\leftrightarrow$ [0,6,0], [$0.5$,10,0]$\leftrightarrow$[$0.5$,12,0], [1,13,0]$\leftrightarrow$[1,15,0],[0,13,0]$\leftrightarrow$[0,13,1], [0,13.5,0]$\leftrightarrow$[0,13.5,1], and [0,14,0]$\leftrightarrow$[0,14,1] (r.l.u)), and no evidence of a magnetic signal was found, disfavoring the possibility of a magnetic structure with non-integer $\vec{k}$ components.
	\begin{figure}[htb]
		\includegraphics[width=\columnwidth]{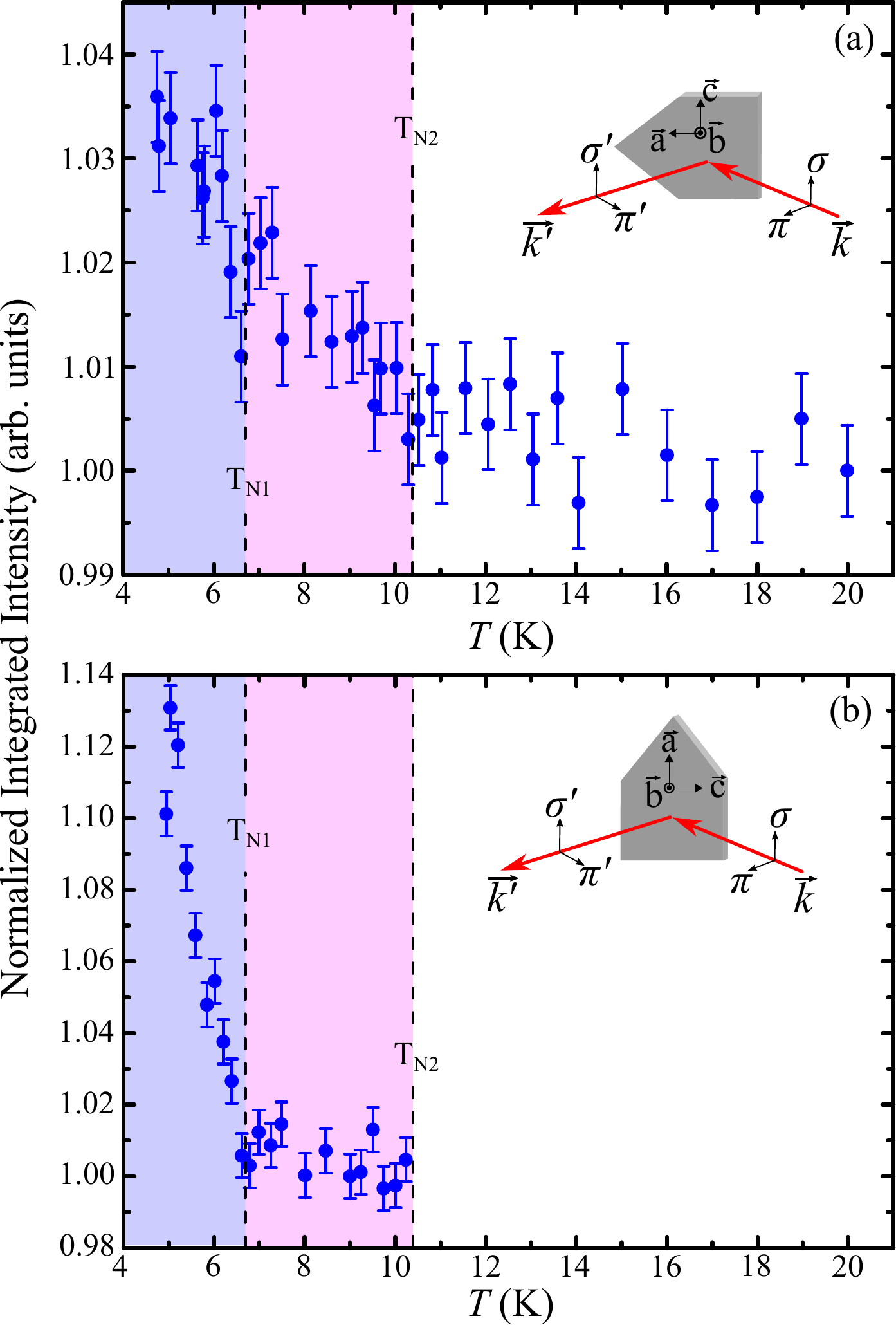}
		\caption{\label{Tdep} Intensities of the (a) $0$ $14$ $0$ reflection in \textit{AB} configuration and (b) $0$ $10$ $0$ reflection in \textit{BC} configuration of HoNiSi$_3$. All the measurements were done on resonance ($E=8.076$ keV) and at the $\pi\sigma$' configuration. The distinct magnetic transition temperatures obtained from these data, $T_{N1} = 6.7(3)$ K (b) and $T_{N2} = 10.3(5)$ K (a), are highlighted. The intensities in (a) and (b) are normalized by the average values above $T_{N2}$ and $T_{N1}$, respectively. Insets: Schematic view of the \textit{AB} and \textit{BC} configurations.}
	\end{figure}
	
\begin{figure}[htb]
		\includegraphics[width=\columnwidth]{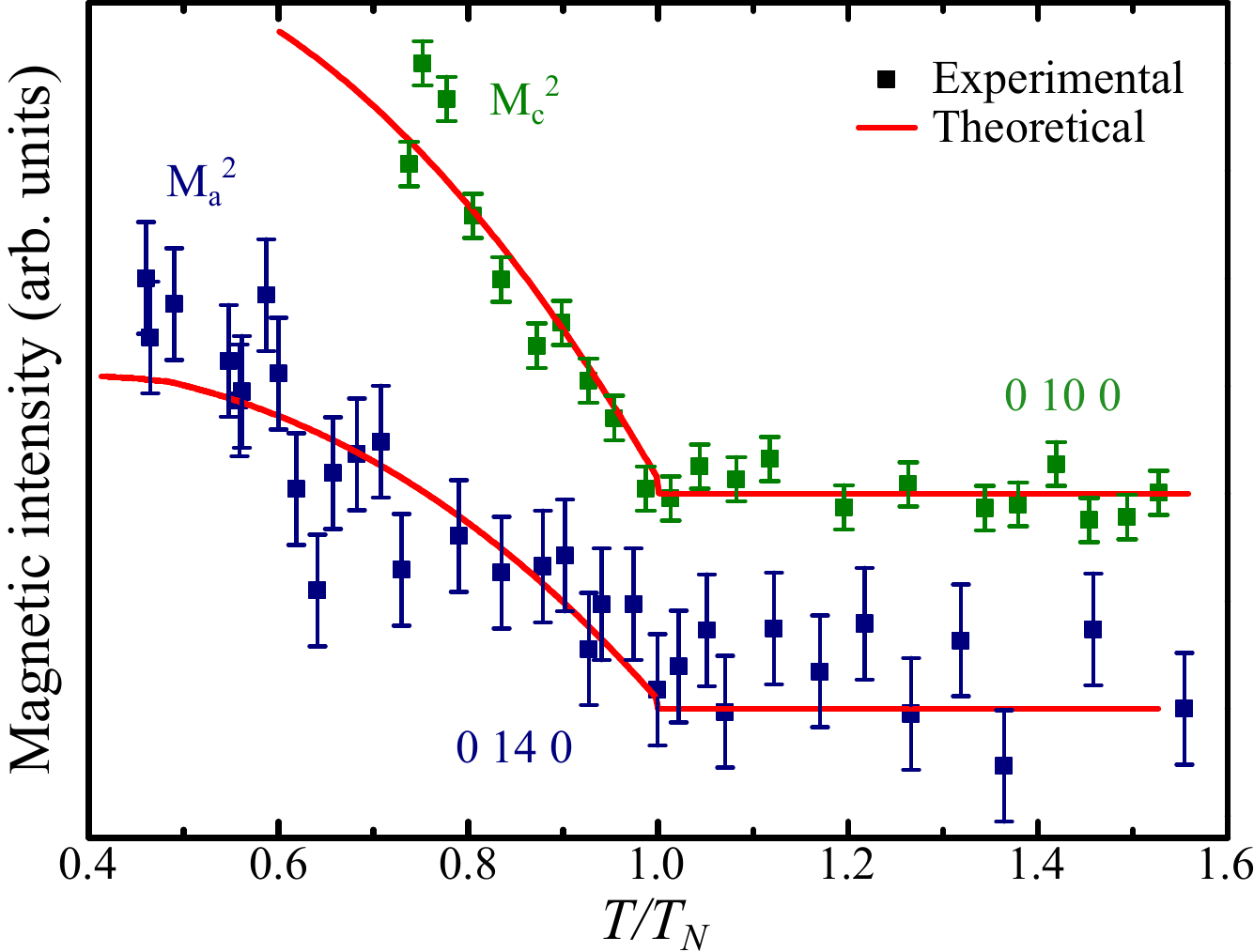}
		\caption{\label{fig:critical} Symbols: Same data of Fig. \ref{Tdep} plotted as a function of the reduced temperature $T/T_N$, where $T_N=T_{N2}=10.3$~K for the 0 14 0 reflection in the \textit{AB} configuration (blue squares) and $T_{N}=T_{N1}=6.7$ K for the 0 10 0 reflection in \textit{BC} configuration (green squares). Data are translated vertically and multiplied by an arbitrary factor for better visualization. Solid red lines are mean-field calculations for the square of the sublattice magnetization along $\vec{a}$ $(M_a^2)$ and along $\vec{c}$ $(M_c^2)$.}
	\end{figure}

	The remaining possibility for the magnetic structure of HoNiSi$_3$ is the same $\uparrow \downarrow \uparrow \downarrow$ stacking along $\vec{b}$ with $\vec{k}~=~[0,0,0]$ found in GdNiSi$_3$ and TbNiSi$_3$ (Ref.~\onlinecite{Tartaglia2019}). This structure retains the $C$ centering of the charge structure, leading to magnetic reflections at the same Bragg positions of the charge reflections. Since magnetic x-ray reflections are dramatically weaker than charge reflections even in \textit{R} $L$-edge resonances, it is a substantial challenge to confirm this magnetic structure. We follow the same methodology employed in our previous work \cite{Tartaglia2019}. Bragg reflections with particularly low structure factors for the charge crystal structure are chosen, and $\pi \sigma$' polarization is employed to further suppress the charge signal, even though some of it is still observed due to polarization leakage. The temperature dependence of the residual intensities is used to evidence any possible magnetic contribution. Figure \ref{Tdep}(a) shows the temperature dependence of the 0 14 0 reflection with the sample mounted in the \textit{AB} configuration, which is sensitive to magnetic moments along $\vec{a}$ (see Sec. \ref{sec:Experiment}). The intensity is nearly constant between $T \sim 11$ and 20 K, whereas a continuous increment is observed below $T_{N2}=10.3(5)$ K, consistent with a magnetic diffraction signal associated with the magnetic ordering transition for $M_a$ previously reported with magnetic susceptibility data \cite{Arantes2018}. Figure \ref{Tdep}(b) shows the temperature dependence of the 0~10~0 reflection intensity in the \textit{BC} configuration, showing a clear increment below $T_{N1} = 6.7(3)$ K that is consistent with the reported magnetic ordering transition temperature for $M_c$ \cite{Arantes2018}. Figure \ref{fig:critical} shows the same experimental data of Fig. \ref{Tdep} plotted as a function of the reduced temperature $T/T_N$, taking as $T_N$ the distinct critical temperatures $T_{N1}$ and $T_{N2}$ for the data taken in the \textit{AB} and \textit{BC} configurations, respectively. This plot is appropriate for comparison with theoretical calculations (see below).
	
\begin{figure}[htb]
		\includegraphics[width=\columnwidth]{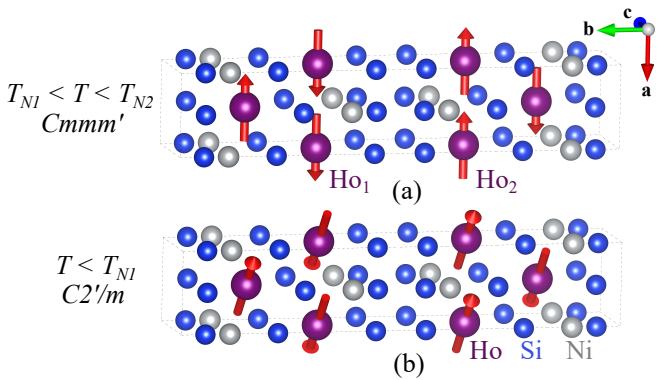}
		\caption{\label{magnetic_structure} Experimental magnetic structure of HoNiSi$_3$ for (a) $T_{N1} < T < T_{N2}$ (phase II) and (b) $T < T_{N1}$ (phase I), determined by a combination of  resonant X-ray magnetic diffraction and macroscopic magnetic experiments, with Ho$_{1}$ and Ho$_{2}$ as used for the symmetry analysis in Sec. \ref{sec:SymmAnls}. Below $T_{N1}$, the magnetization on each site has components both on the $\vec{a}$ and $\vec{c}$ axes.}
	\end{figure}

	Besides confirming the component-separated magnetic transitions in HoNiSi$_3$ by a microscopic technique, our diffraction data reveal a magnetic structure where FM $ac$ planes are stacked in a $\uparrow \downarrow \uparrow \downarrow$ pattern along the $\vec{b}$ direction. The experimental magnetic structures for phase II ($T_{N1}<T<T_{N2}$) and phase I ($T<T_{N1}$) are displayed in Figs. \ref{magnetic_structure}(a) and \ref{magnetic_structure}(b), respectively.
	
	\section{\label{sec:teomet}Theory}
	\subsection{\label{sec:SymmAnls}Symmetry analysis}
	The experimental magnetic structures shown in Figs.~\ref{magnetic_structure}(a) and \ref{magnetic_structure}(b), inferred from the $\uparrow \downarrow \uparrow \downarrow$ stacking pattern obtained here in combination with the moment directions obtained from previous magnetic susceptibility measurements \cite{Arantes2018}, are compared with the symmetry-allowed magnetic structures considering the magnetic propagation vector $\vec{k}~=~[0,0,0]$. Although our diffraction data also suggest magnetic moments along $\vec{a}$ and $\vec{c}$, we cannot rule out from them any component along $\vec{b}$ (not seen in either magnetic susceptibility and magnetization isotherm measurements \cite{Arantes2018}). In the nuclear crystal structure of HoNiSi$_3$ with $Cmmm$ space group, the Ho ions occupy the $4j$ Wyckoff site [$(0,y,0.5)$ and $(0,-y,0.5)$ + $C$-centering atomic coordinates]. The possible magnetic structures were determined independently through representation analysis using the SARAh suite \cite{Wills2000} and the magnetic space group formalism using the Bilbao Crystallographic Server \cite{Perez-Mato2015}. In the decomposition of the magnetic representation, six one-dimensional irreducible representations (\textit{irreps}) of the $Cmmm$ space group, appearing one time each, can generate magnetic ordering. Three of them give rise to FM order, and the remaining ones give rise to AFM structures with moments along each crystallographic direction. These representations are shown in Table \ref{table:irrep} along with their respective magnetic space groups. Thus, at phase II the magnetic structure is described by the $\Gamma_{8}$ (mGM$_{2}^{-}$) representation, or alternatively by the \textit{Cmmm}' magnetic space group. At phase I, an additional component along $\vec{c}$ arises, which can be described with $\Gamma_{4}$ (mGM$_{3}^{-}$). Combining both (mGM$_{2}^{-}$) and (mGM$_{3}^{-}$) representations, the resulting magnetic space group at phase I is $C2$'$/m$.
	
	\begin{table}[htb]
		\caption{Irreducible representations $\Gamma_n$  \cite{Perez-Mato2015} leading to AFM structures with $\vec{k}=[0, 0, 0]$ for the $Cmmm$ space group of the chemical structure, along with the symmetry-allowed magnetic moments at the Ho$_1$ and Ho$_2$ positions not related by the $C$-centering of the crystallographic unit cell (see also Fig.~\ref{magnetic_structure}). The magnetic space group corresponding to each representation is also given.} 
		\centering
		\begin{tabular}{c c c c}
			\hline\hline 
			$\Gamma$& Ho$_{1}$ & Ho$_{2}$ & Magnetic space group\\
			\hline
			$\Gamma_{2}$ (mGM$_{1}^{-}$) & ($0$,$m_{y}$,$0$) & ($0$,-$m_{y}$,$0$) & $C$$m$'$m$'$m$' \\ [0.5ex]
			$\Gamma_{8}$ (mGM$_{2}^{-}$) & ($m_{x}$,$0$,$0$) & (-$m_{x}$,$0$,$0$) & \textit{Cmmm}' \\ [0.5ex]
			$\Gamma_{4}$ (mGM$_{3}^{-}$) & ($0$,$0$,$m_{z}$) & ($0$,$0$,-$m_{z}$) & \textit{Cm}'\textit{mm} \\ [0.5ex] 
			\hline
		\end{tabular}
		\label{table:irrep}
	\end{table}
	
	The possible magnetic structures that fulfill Landau's criteria of second-order phase transitions with a single \textit{irrep} are the ones with magnetic moments pointing along the unit cell directions. In HoNiSi$_{3}$, there are two phase transitions, and for each of them, a single \textit{irrep} drives the transition. The magnetic space group of the highest symmetry that is consistent with these two irreducible representations is monoclinic $C2$'$/m$. Thus, the low symmetry of the magnetic structure below $T_{N1}$ is indicative of a monoclinic lattice, in contrast to the reported orthorhombic $Cmmm$ space group of the charge structure. These considerations point to a symmetry-lowering structural phase transition that occurs simultaneously with the magnetic transition at $T_{N1}$. Such monoclinic distortion was not clearly manifested in our present x-ray diffraction experiment. We should mention that the direct observation of small monoclinic distortions with respect to a parent orthorhombic lattice poses a significant challenge. For such, a high-resolution x-ray diffraction experiment optimized for such goal would be needed, which is beyond the scope of the present paper.
	\subsection{\label{sec:Simu}General structure of the magnetic Hamiltonian}
	The magnetic phases and transitions observed in HoNiSi$_{3}$ can be understood using a magnetic model that considers exchange interactions between the magnetic moments located at the Ho$^{3+}$ ions and CEF effects:
	\begin{equation}
		\mathcal{H}=\mathcal{H}_{int}+\mathcal{H}_{CEF}. \label{eq:H}
	\end{equation}
	In metallic 4\textit{f}-magnetic systems like HoNiSi$_3$ and GdNiSi$_3$, the magnetic couplings are dominated by the RKKY mechanism, which leads to exchange couplings between the magnetic moments at the \textit{R} ions,
	\begin{equation}\label{eq:exchange}
		\mathcal{H}_{int} = \sum_{i < j} K_{ij} \hat{J}_i\cdot \hat{J}_j,
	\end{equation}
	where $\hat{J}_i$ is the angular momentum operator of the magnetic moment located at site $i$ and $K_{ij}$ is the RKKY exchange coupling constant between magnetic moments $i$ and $j$. $K_{ij}$ can be AFM ($K_{ij}>0$) or FM ($K_{ij}<0$) and are expected to decay with the inverse cubic distance between sites $i$ and $j$.
	
	The CEF effects depend on the point symmetry of the \textit{R} ion sites in the lattice and the orbital angular momentum of the ground state multiplet of the ion. The point symmetry of the \textit{R} sites is $C_{2v}$ ($m2m$), which allows for nine CEF terms up to sixth order \cite{walter1984treating}:
	\begin{equation}
		\mathcal{H}_{cef}=\sum_{n = 2,4,6}\sum_{m = 0,2,\ldots,n}B_n^m \hat{\mathbf{O}}_n^m .
		\label{eq:cef}
	\end{equation}
	\subsection{Magnetic simulations guided by density-functional theory calculations}
	\subsubsection{Determination of the exchange couplings\label{exg_coupling}}
	To calculate the exchange coupling parameters, we focus first on the structurally related but simpler material GdNiSi$_3$. In this compound, the magnetic moments at the Gd$^{3+}$ ions are, according to Hund's rule, given by the $L=0$, $S=7/2$ multiplet for which the CEF effects are not relevant \footnote{There can be a small coupling with the $L=1$ multiplet which leads to CEF effects, although much smaller than in HoNiSi$_{3}$}. This makes the density-functional theory (DFT) determination of the total energy global minimum in each magnetic configuration much simpler, avoiding the large uncertainty due to the presence of multiple metastable configurations of the $L\neq 0$ systems \cite{Dorado2009,Larson2007,Amadon2008}. 
	{\it Ab initio} calculations were done following a procedure similar to the one described in Refs. \onlinecite{garcia2021magnetic, facio2015}.
	\begin{figure*}[htb]
		\begin{center}
			\includegraphics[width=2\columnwidth]{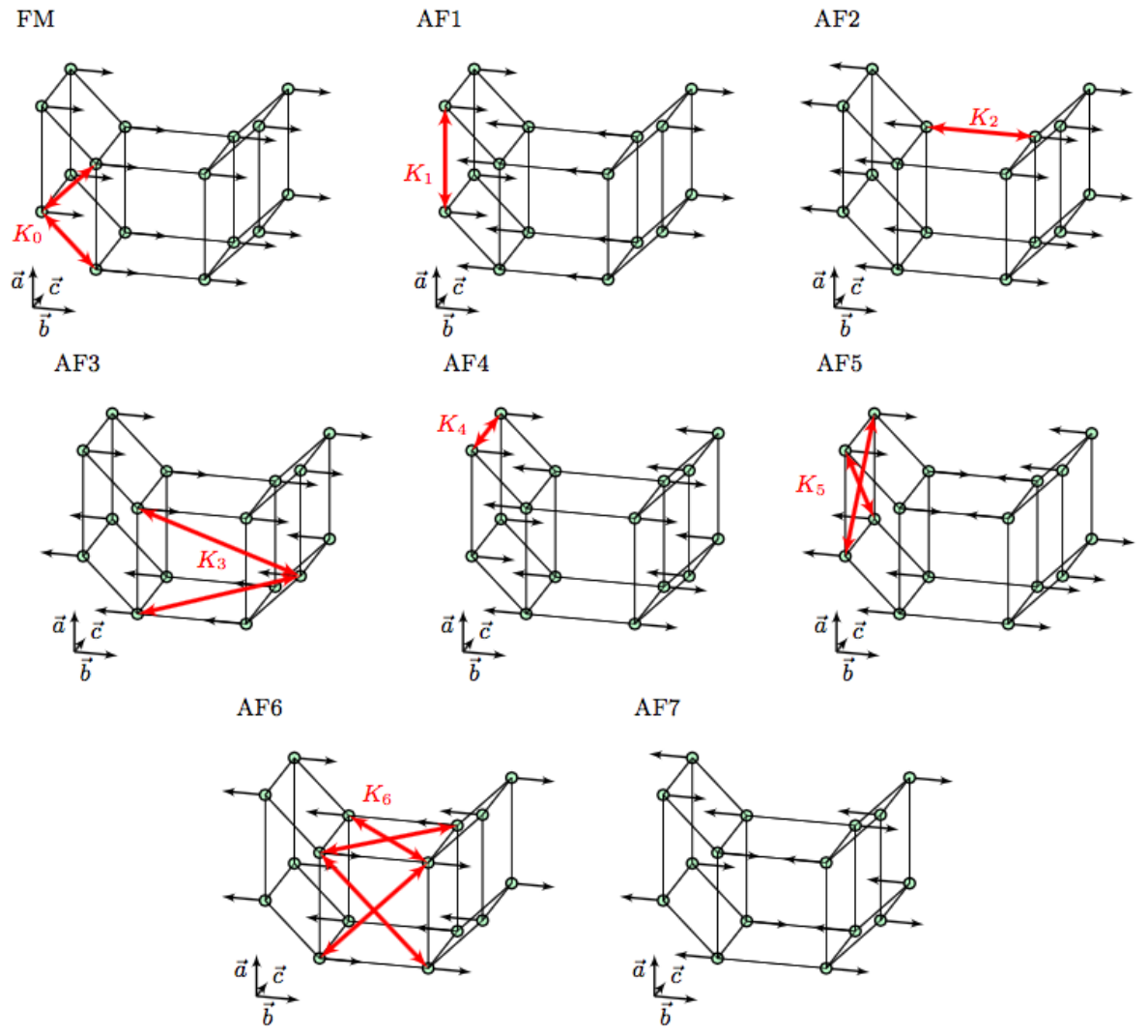}
			\caption{Putative magnetic configurations used for the DFT calculations. The first configuration represents a simple FM order, while the remaining ones are different AFM arrangements. The configuration AF1 is compatible with the magnetic ordering found experimentally for YbNiSi$_3$ \cite{kobayashi2008neutron}, while configuration AF4 corresponds to the magnetic ordering found for Gd/TbNiSi$_3$ \cite{Tartaglia2019}. Each of the seven exchanges $K_i$ is depicted by a red arrow.}
			\label{fig:3conf}
		\end{center}
	\end{figure*}
	\begin{table}[htb]
		\caption{Energy values relative to the AF4 configuration, obtained via DFT calculations for each magnetic configuration shown in Fig. \ref{fig:3conf}. The energies are normalized by $S^2$ and the number of atoms $N$.}
		\centering
		\begin{tabular}{ccc}
			\hline\hline
			Configuration &  & Energy (eV/$S^2$/$N$) \\
			\hline
			FM & & 0.088 \\ [0.5ex]
			AF1 & & 0.011 \\ [0.5ex]
			AF2 & & 0.079 \\ [0.5ex]
			AF3 & & 0.044 \\ [0.5ex]
			AF4 & & 0.000 \\ [0.5ex]
			AF5 & & 0.044 \\ [0.5ex]
			AF6 & & 0.064 \\ [0.5ex]
			AF7 & & 0.045 \\ [0.5ex]
			\hline
		\end{tabular}
		\label{tabener}
	\end{table}
	Total-energy DFT calculations were thus carried out for GdNiSi$_3$ considering eight possible collinear magnetic structures (see Fig. \ref{fig:3conf}).
	These calculations were performed using the generalized gradient approximation (GGA) of Perdew, Burke, and Ernzerhof for the exchange and
	correlation functional as implemented in the Wien2K code \cite{perdew1996generalized,blaha2001wien2k}. A local Coulomb repulsion was included for a better treatment of the highly localized $4f$ states using GGA+\textit{U}, within the fully localized limit for the double counting correction \cite{anisimov1993density}. A value of  $U_{eff}~=~U~-~J~=~6$~eV was used for the effective local Hubbard parameter, which has been successfully implemented before for Gd compounds \cite{betancourth2019magnetostriction}. In the DFT calculations, we considered the experimental lattice parameters \cite{Arantes2018,Tartaglia2019} and relaxed the internal positions. A supercell of $2\times1\times2$ unit cells was used to calculate the exchange couplings out of the magnetic configurations of Fig. \ref{fig:3conf}. In this case, a $9\times3\times9$ $k$-mesh was used to sample the Brillouin zone. The resulting energies are shown in Table \ref{tabener}. The lowest energy was reached for the AF4 structure, which is indeed the experimentally found structure of this compound \cite{Tartaglia2019}. 
	
	The next step is to parametrize the energy of each possible magnetic structure in terms of up to seven exchange coupling parameters $K_i$'s, according to
	\begin{equation}
		\begin{aligned}
			E_{FM}/J^2 &= -2K_0 - 2K_1 - K_2 - 4K_3 \\ & \qquad \qquad \qquad \qquad - 2K_4 - 4K_5 - 4K_6,\\
			E_{AF1}/J^2 &= 2K_0 - 2K_1 - K_2 + 4K_3 \\ & \qquad \qquad \qquad \qquad - 2K_4 - 4K_5 - 4K_6,\\
			E_{AF2}/J^2 &= -2K_0 - 2K_1 + K_2 + 4K_3 \\ & \qquad \qquad \qquad \qquad - 2K_4 - 4K_5 + 4K_6,\\
			E_{AF3}/J^2 &= 2K_1 - K_2 - 2K_4 + 4K_5,\\
			E_{AF4}/J^2 &= 2K_0 - 2K_1 + K_2 - 4K_3 \\ & \qquad \qquad \qquad \qquad - 2K_4 - 4K_5 + 4K_6,\\
			E_{AF5}/J^2 &= 2K_1 + K_2 - 2K_4 + 4K_5,\\    
			E_{AF6}/J^2 &= 2K_0 -2K_1 - K_2 + 4K_3 + 2K_4 + 4K_5,\\ 
			E_{AF7}/J^2 &= 2K_1 + K_2 + 2K_4 - 4K_5 - 4K_6
		\end{aligned}
		\label{eq:energies}
	\end{equation}
	(see Fig. \ref{fig:3conf} for the definition of each $K_i$). By combining the data in Table \ref{tabener} with Eqs. (\ref{eq:energies}), the seven $K_i$'s ($i=0-6$) are directly obtained and shown in Table \ref{tabcts}.
	
	In practice, it is often the case that only a few couplings ($K_{ij}$ for $i$ and $j$ nearest neighbors) need to be considered to obtain an accurate description of the magnetic properties \cite{facio2015,garcia2020magnetic,garcia2021magnetic}. Here, we also consider a simplified model with only three independent exchange constants, namely $K_0$, $K_1 \equiv K_4$, and $K_2$, therefore setting $K_3=K_5=K_6=0$ (see Fig. \ref{fig:3conf}). The constrained exchange constants are obtained by the procedure described above, and the results are also shown in Table~\ref{tabcts}.
	
	\begin{table}[htb]
		\caption{Exchange $K_i$ constants values obtained through DFT and mean-field fittings for GdNiSi$_3$. For Ho, only the de Gennes scaling values from mean-field fittings are presented. Positive(negative) exchange represents AFM (FM) interactions.}
		\centering
		\begin{tabular}{cccc}
			\hline\hline
			& DFT & \multicolumn{2}{c}{Simplified mean-field} \\
			\hline
			$K_i$ constant   & Gd (K)    & Gd (K)  & Ho (K) \\
			\hline
			$K_0$       &2.31    &1.86      &0.10 \\ [0.5ex]
			$K_1$       &0.93    &-0.214    &-0.012 \\ [0.5ex]
			$K_2$       &-0.030  &1.16      &0.064 \\ [0.5ex]
			$K_3$       &-0.020  &        	& \\ [0.5ex]
			$K_4$       &-0.78   &-0.214    &-0.012 \\ [0.5ex]
			$K_5$       &-0.46   &	        & \\ [0.5ex]
			$K_6$       &0.15    &	        & \\ [0.5ex]
			\hline
		\end{tabular}
		\label{tabcts}
	\end{table}
	
	Once the magnetic exchange couplings for GdNiSi$_3$ are obtained, the corresponding ones for HoNiSi$_3$ can be estimated using a de Gennes scaling \cite{blundell2003magnetism}. This scaling, usually valid for most \textit{R}, considers that the interactions between magnetic moments only involve the spin part of the total magnetic moment. Under this hypothesis, the couplings can be re-scaled, projecting the spin moment onto the total magnetic moment, resulting in $K_{ij}($\textit{R}$)=(g_J-1)^2 K_{ij}($Gd$)$ (the square comes from the two-moment interaction that involves two projections), where $g_J$ is the gyromagnetic factor of the \textit{R} being considered. For \textit{R} compounds, this scaling is frequently performed to estimate the ordering temperature \cite{Arantes2018, mercena2021crystalline}. The thus obtained $K_i$ values for HoNiSi$_3$ under the simplified model with three independent exchange constants are also given in Table \ref{tabcts}.
	
	Classical (CMC) and Quantum Monte Carlo (QMC) simulations using the ALPS package \cite{bauer2011alps,albuquerque2007alps}, as well as mean-field calculations using $\mathcal{H}_{int}$ were performed to obtain the magnetization and specific-heat curves for GdNiSi$_3$. The results for $T_N$ and the Curie Weiss (CW) temperature $\theta$ are shown in Table \ref{tab:tn}. 
	The advantage of using the mean-field model is the possibility of finding an analytic expression for $T_N$ and $\theta$ as a function of the couplings. For this system, they are given by
	\begin{subequations}
		\begin{align}
			T_N & = \frac{1}{3}  J (J+1) (2 K_0 - 2 K_1 + K_2 - 4 K_3 \nonumber \\
			& \qquad  \qquad \qquad - 2 K_4 - 4 K_5 + 4 K_6 ), \label{eqTN}\\
			\theta & = \frac{1}{3} J (1+J) (2 K_0 + 2 K_1 + K_2 + 4 K_3  \nonumber \\
			& \qquad  \qquad \qquad + 2 K_4 + 4 K_5 + 4 K_6  ).\label{eqTheta}
		\end{align}
	\end{subequations}
	If we consider the simplified model, these equations yield $T_N$ and $\theta$ shown in Table \ref{tab:tn}. Considering the full model (seven exchange couplings), the values are slightly modified to $T_N=36$ K and $\theta=-19$ K.
	
	Although the mean-field approximation leads to an overestimation of the transition temperature of GdNiSi$_3$, it provides the correct physical picture. We thus base our analysis of HoNiSi$_3$ on the mean-field approximation to be able to perform simulations for a wide range of the model's parameters.
	
	\begin{table}[htb]
		\caption{$T_N$ and $\theta$ of GdNiSi$_3$ obtained from experimental data, CMC, QMC, and mean-field models using the DFT parameters shown in Table \ref{tabcts}.}
		\centering
		\begin{tabular}{c c c c c c}
			\hline\hline
			& \multirow{2}{*}{Exp.} & \multirow{2}{*}{CMC} & \multirow{2}{*}{QMC} & \multicolumn{2}{c}{Mean-field} \\
			& & & & Full & Simplified \\  
			\hline
			$T_N$ (K) & 22.2(2) & 13.5(5) & 17.5(5)& 36 & 30.1 \\ [0.5ex]
			$\theta$ (K) & -30(3) & -19.9(3) &  -27.2(3)& -19 & -21.1 \\ [0.5ex]
			\hline
		\end{tabular}
		\label{table:tabmodels}
		\label{tab:tn}
	\end{table}
	
	\subsubsection{Minimal model and mean-field approximation for HoNiSi$_{3}$\label{MeanFieldResults}}
	
	In this section, we develop a minimal model that is able to explain the available experimental results.
	As we show below, the bulk properties of HoNiSi$_3$ can be explained using only two of the nine CEF terms in Eq.~(\ref{eq:cef}). The tendency of the magnetic moments to stay in the $ac$ plane, as observed in both low-temperature AFM phases, can be accounted for using the CEF term, 
	\begin{equation}
		\hat{O}_2^2= J_a^2-J_b^2,
		\label{eq:O22}
	\end{equation}
	with a negative coefficient $B_2^2$.
	The observed tilting of the magnetic moment towards the $\vec{c}$ direction in phase I can be described using a positive $B_4^0$ for the CEF operator: 
	\begin{multline}
		\hat{O}_4^0 = 35J_c^4 - 30J(J+1)J_c^2 +25J_c^2 \\ + 3J^2(J+1)^2 - 6J(J+1).
		\label{eq:O40}
	\end{multline}
	
	\begin{figure}[htb!]
		\includegraphics[width=0.48 \textwidth]{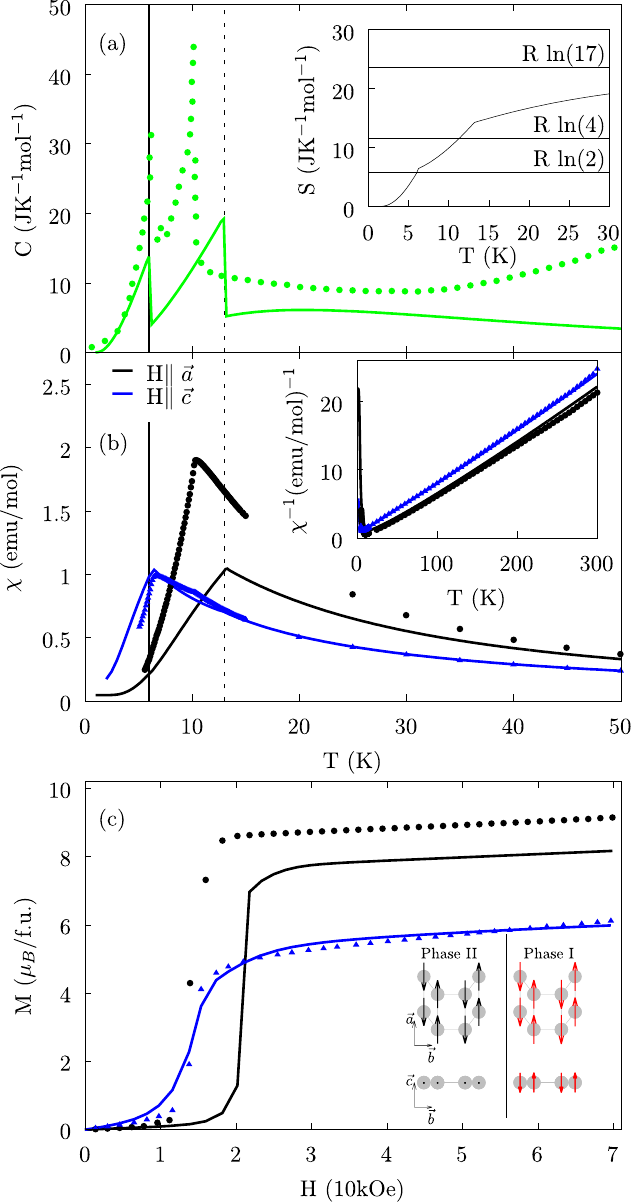}
		\caption{\label{fig:MFHo} Mean-field results (lines) and experimental data \cite{Arantes2018} (symbols) for (a) total magnetic specific heat of HoNiSi$_3$ as a function of the temperature. Inset shows the released entropy up to 30 K. (b) The magnetic susceptibility as a function of temperature for external magnetic field of $H = 1$ kOe along with $\vec{a}$ and $\vec{c}$ directions. Inset shows the inverse magnetic susceptibility up to 300 K (lines are almost indistinguishable from symbols). (c) Magnetization at $T = 2.2$ K as a function of an external magnetic field parallel to the $\vec{a}$ and $\vec{c}$ directions. Inset shows the mean-field magnetic structures at phases I and II.}
	\end{figure}	
	
	Mean-field \footnote{A detailed explanation of the application of the mean-field method used here for magnetic RE can be found in Ref. \cite{jensen1991rare}.} calculations of the magnetization as a function of temperature and external magnetic field were performed to fit the available experimental data \cite{Arantes2018} using $B_2^2$ and $B_4^0$ as fitting parameters. Including all CEF parameters allowed by symmetry up to the sixth order in the fitting procedure does not lead to a significant improvement of the fits, nor does it change the physical picture obtained using only two parameters (nevertheless, their effect is examined in Sec. \ref{ExploringUMAP}). In the fitting procedure of the simplified model presented in this subsection, a smaller de Gennes factor ($\sim$ 0.055 rather than 0.0625) for Ho$^{3+}$ was used to compensate for the overestimation of the transition temperatures by the mean-field approximation. The obtained parameters are $B_2^2 = -0.85$ K and $B_4^0 = 2.1$ mK. 
	
	Figure \ref{fig:MFHo} presents the mean-field results (solid lines) for the simplified magnetic Hamiltonian using the estimated model parameters obtained as described above. For a comprehensive comparison, we included the experimental data (filled symbols) taken from Ref. \onlinecite{Arantes2018}. First, two peaks in the magnetic specific heat as a function of temperature emerge in the mean-field results, corresponding to a paramagnetic (PM)-AFM and an AFM-AFM transition. Additionally, as can be seen in Fig.~\ref{fig:MFHo}(a), the transition temperatures are in good agreement with the measured values. The magnetic entropy [see inset of Fig. \ref{fig:MFHo}(a)] is $\sim R \ln(4)$ for $T\sim T_{N2}$, which can be attributed to the $B_4^0$ CEF term (see below). Also, the magnetic susceptibility in the $\vec{a}$ direction as a function of the temperature presents a peak at $T_{N2}$, while the corresponding one in the $\vec{c}$ direction has a peak at $T_{N1}$ [see Fig. \ref{fig:MFHo}(b)], in close similarity to the experimental results. The behavior of $\chi_{c}$ follows a CW law for $T>T_{N1}$ while $\chi_{a}$ follows closely a CW law for $T>T_{N2}$. Figure \ref{fig:MFHo}(c) shows the calculated magnetization as a function of the external magnetic field. As in the experimental data, it presents a flop transition to a state where the magnetic moments in the $\vec{a}$ or $\vec{c}$ direction become FM for an external magnetic field in the same direction. We also see that for both theoretical and experimental results, the higher magnetization is attained when the magnetic field is along the $\vec{a}$ direction. The resulting magnetic structure from the model is depicted in the inset of Fig. \ref{fig:MFHo}(c), in full agreement with the experimental structure determined in this work (see Sec. \ref{sec:MagDiff}).
	
	As the intensities of the AFM Bragg reflections reported in Sec. \ref{sec:MagDiff} are proportional to the square of the sublattice magnetization \cite{Wills2001}, they can be also calculated using the mean-field model. The solid lines in Fig. \ref{fig:critical} show $M_a^2$ and $M_c^2$ as a function of the reduced temperature $T/T_N$, where $T_N$ is taken here as the mean-field $T_{N1}$ for the $M_c^2$ curve and $T_{N2}$ for $M_a^2$. It can be seen that the comparison with experimental data is quite satisfactory.
	
	\begin{table*}[htb]
		\centering
		\caption{\label{autoestados} Eigenvalues $E_{i}$ (in K) and  associated eigenfunctions of Ho$^{3+}$ ($J$ = 8) for the CEF Hamiltonian  $\mathcal{H}_{cef}$ with parameters $B_2^2 = -0.85$ K and $B_4^0 = 2.1$ mK.}
		\begin{tabular}{c c c}
			\hline\hline
			$E_{i}$ & &Eigenfunctions \\ 
			\hline
			0.0 & &$ 0.46 (\estado{-6}  +  \estado{-4} + \estado{4}  +  \estado{6}) + 0.24 ( \estado{-2} + \estado{2})  + 0.17 \estado{0} $  \\ [0.5ex]
			0.2 & &$ 0.19 (\estado{-7} +  \estado{-1} + \estado{1} +  \estado{7}) + 0.55 (\estado{-5} + \estado{5})  + 0.35 ( \estado{-3}     + \estado{3} )  $ \\ [0.5ex] 
			1.8 & &$0.51 (\estado{-6}- \estado{6}) +0.46 (\estado{-4}-\estado{4}) +0.18 (\estado{-2}  - \estado{2} )    $ \\ [0.5ex]
			2.2 & &$ 0.21 (\estado{-7}-\estado{7})  + 0.59 (\estado{-5} -\estado{5})  + 0.32 (\estado{-3} -\estado{3})$ \\ [0.5ex]
			27.6 & &$0.47 (\estado{-6}+\estado{6})  - 0.19 (\estado{-4}+\estado{4})  - 0.40 (\estado{-2}  + \estado{0}  +  \estado{2} )  $ \\ [0.5ex]
			31.9 & &$0.33( \estado{-7} -\estado{-3}-\estado{3}+\estado{7}) +0.26( \estado{-5} +\estado{5})    - 0.46 (\estado{-1}  + \estado{1} ) $ \\ [0.5ex]
			36.4 & &$ 0.48 ( -\estado{-6} +\estado{6} ) +0.41 (\estado{-4}-\estado{4}) +0.32 (\estado{-2}  - \estado{2} )  $ \\ [0.5ex]
			43.1 & &$0.57 (-\estado{-7} + \estado{7})  + 0.39 ( \estado{-3} -\estado{3}) + 0.16 (\estado{-1} -\estado{1} )   $ \\ [0.5ex]
			51.5 & &$0.59 (\estado{-7} +\estado{7} - 0.30 (\estado{-5} +\estado{5} ) +0.25 (\estado{-1} +\estado{1}) $ \\ [0.5ex]
			58.4 & &$0.24 (\estado{-6}+\estado{6})  - 0.48 (\estado{-4}+\estado{4}) +0.26 (\estado{-2}+\estado{2}) +0.53 \estado{0}  $ \\ [0.5ex]
			59.9 & &$0.37 (\estado{-7}-\estado{7}) +0.39( -\estado{-5}+\estado{5})  + 0.41 (\estado{-3} - \estado{3})  + 0.22 (\estado{-1} - \estado{1} ) $ \\ [0.5ex] 
			86.1 & &$0.35 (-\estado{-4}+\estado{4}) +0.61 (\estado{-2}  - \estado{2} ) $ \\ [0.5ex]
			86.7 & &$0.20 (\estado{-5}+\estado{5}) -0.51 ( \estado{-3}+\estado{3})  + 0.44 (\estado{-1} + \estado{1} ) $ \\ [0.5ex]
			103.3 & &$0.70( \estado{-8} + \estado{8} ) $ \\ [0.5ex]
			103.3 & &$0.70 (\estado{-8}  -  \estado{8} ) $ \\ [0.5ex]
			126.1 & &$0.28(- \estado{-3} +\estado{3}) +0.65 (\estado{-1}  - \estado{1} ) $ \\ [0.5ex]
			126.2 & &$0.14 (\estado{-4} +\estado{4})  - 0.46 (\estado{-2} + \estado{2}) + 0.73 \estado{0}  $ \\ [0.5ex]
			\hline
		\end{tabular}
	\end{table*}
	
	To gain further insight into the physical origin of the observed AFM-AFM transition, we also analyze the system under the molecular field approximation. In the mean-field approach, a cluster of eight Ho$^{3+}$ ions was used to determine the magnetic order as a function of temperature and external magnetic field. In the absence of an external magnetic field, the two ordered phases correspond to the AF4 configuration, differing only in the direction of the magnetic moments. As a consequence, the mean-field approach can be reduced to a molecular field approximation in which a single magnetic moment is under the influence of the CEF and of an effective magnetic field generated by the interaction with the other magnetic moments:
	\begin{equation}
		H_{mol} = B_2^2 \hat{O}_2^2+B_4^0 \hat{O}_4^0  - \vec{H}_{eff}\cdot \hat{J}.
		\label{eq:mo}
	\end{equation}
	Here $\vec{H}_{eff}=\lambda \langle \hat{J}\rangle$, where $\lambda$ is determined by the exchange interactions, and $\langle \hat{J}\rangle$ is calculated in a self-consistent way. 
	For a single magnetic moment with $J=8$, the Hilbert space is spanned into $2J+1$ states ($\vert m \rangle$ with $m=-J, \dotsm,  J$). The eigenvalues and eigenvectors of $H_{mol}$ can be readily obtained by diagonalizing the associated $17\times 17$ matrix. This allowed us to obtain $\langle \hat{J}\rangle$ at finite temperatures and find a self-consistent solution. A numerical calculation pursuing this route reproduces the mean-field results once the correct AF4 order is selected to determine $\lambda$. 
	
	In the PM phase, $H_{eff}=0$, and $H_{mol}$ is reduced to the CEF terms. For simplicity, we set at this point $B_2^2$ to zero, but we reintroduce it at a later stage. The remaining term $\hat{O}_4^0$ with a positive $B_4^0$ gives rise to a fourfold ground state degeneracy ($J_c = -6, -5, 5, 6$) which is consistent with the entropy [$\sim R\ln(4)$] obtained in the PM phase for $T\gtrsim T_{N2}$ [see inset of Fig. \ref{fig:MFHo}(a)]. 
	
	For temperatures slightly below $T_{N2}$, a non-zero $\vec{H}_{eff}$ emerges, signaling the transition to the AFM phase. The direction of $\vec{H}_{eff}$ is given by the direction of maximal magnetic susceptibility and determines the direction of $\langle \hat{J}\rangle$. To find the direction of maximal susceptibility, we turn on a small external magnetic field ($H\ll T_{N2}$) in the PM phase ($T \gtrsim T_{N2}$), where $\langle \hat{J}\rangle=0$, and consider the $\vec{a}$ and $\vec{c}$ directions (in the absence of the $\hat{O}_2^2$  term the problem is symmetric under rotations around the $\vec{c}$ axis). An external magnetic field in the $\vec{c}$ direction does not change the eigenvectors of the system ($J_c$ is a good quantum number for $B_2^2=0$) but changes their relative energies, leading to a susceptibility proportional to $1/T$. A magnetic field in the $\vec{a}$ direction, however, produces a different effect because $J_c$ is no longer a good quantum number. The magnetic field mixes terms that differ in $\Delta J_c=\pm 1$ and leads to a susceptibility in the $\vec{a}$ direction that does not decrease as the temperature increases up to sufficiently high temperatures where it becomes larger than the one in the $\vec{c}$ direction. The PM to AFM transition occurs at a temperature where $\chi_a>\chi_c$.
	
	The inclusion of the $\hat{O}_2^2$ term using the estimated value for $B_2^2 = -0.85$ K leads to a small breaking of the ground state degeneracy (the energies of the four lowest lying states differ by $\sim 2$ K, see Table \ref{autoestados}), but does not change the entropy significantly for $T\sim T_{N2}$. This term further increases the magnetic susceptibility in the $\vec{a}$ direction compared to $\vec{c}$ and reduces the temperature above which the susceptibility in the $\vec{a}$ direction becomes larger than in the $\vec{c}$ direction. It also breaks the symmetry between the $\vec{a}$ and $\vec{b}$ directions, decreasing the magnetic susceptibility in the latter direction.
	
	Below $T_{N2}$, the magnetic moments order in the $\vec{a}$ direction. As a result, the susceptibility with the field in this direction decreases while the susceptibility with the field in the $\vec{c}$ direction keeps increasing [see Fig. \ref{fig:MFHo}(b)].
	At sufficiently low temperatures, the susceptibility in the $\vec{a}$ direction is no longer the largest, and it becomes energetically favorable to tilt $\langle \hat{J}\rangle$, with a component in the $\vec{c}$ direction. This leads to the AFM-AFM transition at $T_{N1}$. The tilting angle can be obtained considering a classical magnetic moment and minimizing the energy of the CEF. At low temperatures, the magnetic moment is contained in the $ac$ plane and forms an angle $\alpha$ with the $\vec{c}$ axis, where
	
	\begin{equation}
		\tan(\alpha) = \sqrt{\frac{5 B_4^0 (8 J^2- 6 J+5 )+B_2^2}{5 B_4^0 (6 J^2+ 6 J-5 )-B_2^2}}. 
		\label{eq:angle} 
	\end{equation}
	
	Using $B_2^2 = -0.85$ K and $B_4^0 = 2.1$ mK estimated above and $J=8$, we obtain $\alpha=41.2^{\circ}$. Future microscopic experiments, such as (i) neutron diffraction when larger crystals become available, (ii) nuclear magnetic resonance \cite{Shioda2021,Ihara2023}, or (iii) resonant x-ray magnetic diffraction experiment with a more efficient rejection of charge scattering and a geometry allowing for azimuthal scans, may be able to determine $\alpha$ experimentally, which could then be compared with our predicted value.
	
	\subsection{Full model ground-state configurations} \label{ExploringUMAP}
	
	In this subsection, we explore the parameter space of the magnetic model [see Eq. (\ref{eq:H})] and the corresponding ground-state configurations consistent with the existing magnetization and magnetic susceptibility data. Our objective is to use HoNiSi$_3$ as a case study to analyze whether the possible magnetic structures can be constrained using this experimental data. Restricting the possible ground states may assist in directing subsequent experiments and DFT calculations to accurately determine the magnetic structure of a material. Accordingly, we exclude here the x-ray data and coupling parameters derived from DFT plus de Gennes scaling in this analysis.
	
	To perform this study, we focus on magnetic structures with an eight-site cluster that corresponds to two crystallographic unit cells (repeated in the $a$ direction). Our proposed magnetic model incorporates the nine CEF terms permitted by symmetry and the seven exchange couplings illustrated in Fig. \ref{fig:3conf}. The model parameters could theoretically be determined by fitting the experimental data to the outcomes from solving the model Hamiltonian. However, due to the high number of parameters, limitations inherent in the model, and the computational demands of solving it via Monte Carlo methods, this strategy is deemed impractical.
	
	Instead, we adopt an approximate mean-field approach, which generally offers a rapid and reliable means to capture the primary qualitative aspects of the experimental data for compounds exhibiting magnetic order. The primary limitation of this approach is its inability to uniquely determine the parameters, as multiple parameter sets may yield similar fit qualities.
	
	To address this challenge, we explored the parameter space beginning with randomly chosen parameters, employing a subplex minimization technique \cite{rowan1990functional} to optimize the fit. The minimized cost function is defined as
	\begin{widetext}
		\begin{equation}
			\Delta^2 = \sum_{\zeta=a,c} \left( \beta_{\zeta} \sum_{i} (\chi_{\zeta,T_i}^{Exp} -\chi_{\zeta,T_i}^{Teor})^2 + \gamma_{\zeta} \sum_{i} (M_{\zeta,H_i}^{Exp} - M_{\zeta,H_i}^{Teor})^2 \right).
		\end{equation}
	\end{widetext}
	Here $\chi_{\zeta,T}^{Exp}$ and $M_{\zeta,H}^{Exp}$ represent the experimental magnetic susceptibility and magnetization, at temperature $T$ and field $H$, respectively, and $\zeta={a,c}$ indexes the external field direction. The theoretical mean-field values are denoted by $\chi_{\zeta,T}^{Teor}$ and $M_{ \zeta,H}^{Teor}$. The normalization factors $\beta_{\zeta}$ and $\gamma_{\zeta}$ are determined through a preliminary minimization process to ensure a balanced contribution from both magnetic susceptibility and magnetization to the cost function.  The minimization process is repeated for 1000 random initial parameter sets and the 200 fits with the lowest cost function are selected. Finally, the ground states corresponding to the selected sets of parameters are classified using a machine learning approach. 
	
	To characterize ground-state magnetic structures to be fed to the machine-learning procedure, we use the square modulus of the spin structure factor,
	\begin{equation}
		S_{\vec{Q}}^{\zeta}  = \left\vert \sum_l \langle J_{\vec{R}_l}^{\zeta} \rangle e^{i \vec{R}_l \cdot \vec{Q} } \right\vert^2,
	\end{equation}
	where $\langle J_l^{\zeta} \rangle$ is the mean value of the  $\zeta=a,b,c$ component of the magnetic moment at site $\vec{R}_l$, and $\vec{Q}=(\pi n_a / a)\hat{a} + (2 \pi n_b / b)\hat{b}$, where $a$ and $b$ are the lattice parameters of the conventional cell of HoNiSi$_3$, $n_a=0,1$, and $n_b=0,1,2,3$. $S_{\vec{Q}}^{\zeta}$ is insensitive to symmetry-related configurations (e.g., an inversion of all magnetic moments). 
	
	The $24=3\times 2\times 4$ values of $\vec{Q}$ form the {\it feature vector} $\vec{v}=\left(S^a_{0,0},S^a_{0,1},\ldots S^c_{1,3}\right)$ for the machine learning analysis, where $n_a$ and $n_b$  in $S^\zeta_{n_a,n_b}$ determine the $\vec{Q}$ value. 
	The similarity between the ground states is quantified by the Euclidean distance between the different feature vectors. 
	
\begin{figure}[htb]
		\includegraphics[width=\columnwidth]{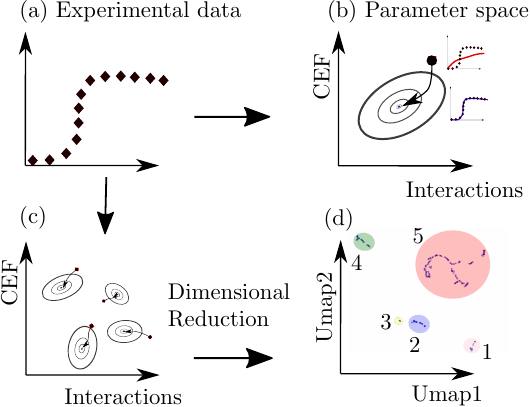}
		\caption{\label{fig:exploracion} Flowchart of the procedure used to catalog the possible fitting parameters. (a) shows a general measurement that can be described by a Hamiltonian with an interaction and a CEF term. (b) shows that in the multidimensional parameter space, a regular seed can be good enough to find an excellent fit (continuous lines represent fit quality contours). (c) illustrates the case for HoNiSi$_3$: Most seeds fall on different fitting parameters. An absolute minimum could be hidden on the irregular landscape. (d) shows the 2D  projection of the different feature vectors originally embedded in the 24-dimensional space after applying UMAP. In this panel, the feature vector (of each dot) corresponds to the 24 components of the square modulus of the spin structure (see main text). The obtained clusters are clearly separated. Figure \ref{fig:clusters} shows explicitly the spin structures (configuration) that are grouped in each cluster.
		}
	\end{figure}

	To analyze the data, we use the Uniform Manifold Approximation and Projection (UMAP) \cite{umap} procedure, a dimension reduction algorithm (as implemented in Tensorflow \cite{tf}).  UMAP is a state-of-the-art unsupervised machine learning algorithm for dimension reduction based on manifold learning techniques and topological data analysis. It works by estimating the topology of high-dimensional data and using this information to construct a low-dimensional representation that preserves the proximity relationships in the data. This dimensional reduction is useful for visualizing the data and for clustering. The steps of this procedure are described schematically in Fig. \ref{fig:exploracion}. Each dot in Fig. \ref{fig:exploracion}(d) represents a $2$D projection of the original feature vectors $\vec{v}$ (we recall that  $\vec{v}$ is the square modulus of the structure factor). Five clusters corresponding to five different ground-state magnetic structures can be clearly distinguished.
	
\begin{figure}
		\includegraphics[width=\columnwidth]{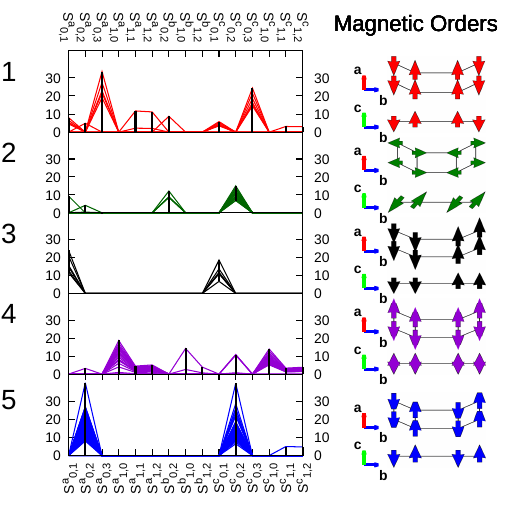}
		\caption{\label{fig:clusters} Clustered square modulus of the structure factor and representative magnetic configurations. The cluster labels correspond to the ones determined in Fig. \ref{fig:exploracion}(d). The plot on the left shows the non-zero components of the square modulus of the structure factor $S_{n_a,n_b}^{\zeta}$ for all the selected ground states. The panel on the right shows two projections of a representative spin configuration for each cluster.}
	\end{figure}

	The $S^\zeta_{n_a,n_b}$ values corresponding to the magnetic configurations for all ground states in a given cluster are shown in Fig. \ref{fig:clusters}, where only the 15 non-zero components of the square modulus of the structure factor  are plotted for each $\vec{v}$. Nine of the $S^\zeta_{n_a,n_b}$ are zero for all ground states. For example, there are no configurations with weight on $S_{\vec{Q}=(0,0)}$ that would correspond to a FM component. Cluster 5 has correlations that correspond to those observed in GdNiSi$_3$, TbNiSi$_{3}$ (magnetic order AF4 in Fig. \ref{fig:3conf}), as well as in HoNiSi$_3$. The best fit obtained (the lowest value of $\Delta^2$) corresponds to a magnetic configuration found within this cluster. The ground-state configurations in cluster 4 present AFM (FM) correlations along the $\vec{a}$($\vec{b}$) axis. Cluster 3 corresponds to states with FM planes stacked antiferromagnetically in a $\uparrow \uparrow \downarrow \downarrow$ pattern while in cluster 1 the stacking pattern is $\uparrow \downarrow \downarrow \uparrow$. In the latter, correlations along the $\vec{b}$ axis are similar to the ones observed in YbNiSi$_3$. Finally, in cluster 2, the correlations are similar to those found in cluster 5 but with a non zero spin component in $\vec{b}$ direction.
	
	The ground-state configurations in cluster 5 correspond to the one obtained experimentally and deduced from the DFT analysis (Sec. \ref{MeanFieldResults}). See Table \ref{table:anothersetofparameters} for a set of parameters in cluster 5 that yields a fit to magnetic measurement as good as the one reported in Sec. \ref{MeanFieldResults}. 
	
	\begin{table*}
		\centering
		\caption{\label{table:anothersetofparameters} Two sets of parameters. Those in the simple model column are the ones used in Sec. \ref{MeanFieldResults}. Those in the optimized column are obtained by a minimization procedure as described in Sec. \ref{ExploringUMAP}. For comparison, we show in the last line the value of the cost function $\Delta^2$ for the two sets of parameters.}
		\begin{tabular}{c c c}
			\hline\hline
			Parameter  & Simple model & Optimized \\ 
			\hline
			$K_0$       &0.10    & 0.066  \\ [0.5ex]
			$K_1$       &-0.012  & 0.014  \\ [0.5ex]
			$K_2$       &0.064   & -0.00018   \\ [0.5ex]
			$K_3$       &      & -0.012    \\ [0.5ex]
			$K_4$       &-0.012   & -0.011    \\ [0.5ex]
			$K_5$       &       & -0.018   \\ [0.5ex]
			$K_6$       &       & 0.034   \\ [0.5ex]
			$O_2^0$     &       & 0.16   \\ [0.5ex]
			$O_2^2$     &-0.85    & -0.7   \\ [0.5ex]
			$O_4^0$     &0.0021   & 0.0083   \\ [0.5ex]
			$O_4^2$     &       & -0.026   \\ [0.5ex]
			$O_4^4$     &       & -0.0055   \\ [0.5ex]
			$O_6^0$     &       & -0.000052   \\ [0.5ex]
			$O_6^2$     &       & -0.000093   \\ [0.5ex]
			$O_6^4$     &       & 0.00013  \\ [0.5ex]
			$O_6^6$     &       & -0.00028   \\ [0.5ex]
			$\Delta^2$  &225     & 204   \\ [0.5ex]
			
			\hline
		\end{tabular}
		
	\end{table*}
	
	This analysis shows that in spite of the complexity of this system due to its low symmetry, giving rise to a large set of CEF and coupling parameters, the magnetic susceptibility and magnetization experimental data can be used to narrow considerably the search for possible ground states using computationally inexpensive mean-field calculations and basic machine-learning tools.
	
	\section{\label{sec:Conclusione}Conclusions}
	In summary, resonant x-ray diffraction experiments were conducted on HoNiSi$_{3}$ in the temperature range where distinct magnetically ordered phases I and II were inferred from previous specific heat and magnetic susceptibility measurements \cite{Arantes2018}. Our presented data show that both phases are characterized by a commensurate magnetic structure with propagation vector $\vec{k} = [0,0,0]$ formed by a $\uparrow\downarrow\uparrow\downarrow$ stacking pattern of FM $ac$ planes with Ho magnetic moments being parallel to $\vec{a}$ axis in phase II and within the $ac$ plane in phase I. A symmetry analysis indicates that the magnetic phase I is not consistent with the presumed $Cmmm$ space-group symmetry of the chemical crystal structure, and therefore a (possibly very small) monoclinic distortion is inferred. Magnetic simulations were performed using different approaches to guide the choice of exchange and CEF parameters. First, a simplified model using a reduced number of fixed exchange parameters obtained from DFT and a few CEF terms taken as fitting parameters was able to capture the experimental magnetic structure, as well as the magnetic susceptibility, magnetization, and specific heat curves. In addition, a methodology based on an unsupervised machine learning-algorithm was employed to search for the possible magnetic structures of the ground state. Remarkably, the parameters that give the best comparisons to the experimental susceptibility and magnetization data, as well as those that are consistent with the simplified model, belong to the same cluster that yields the correct magnetic structure. The methodology employed here may be extended to other magnetic materials where the complete set of exchange and CEF parameters are not known {\it a priori}. 
	
	\begin{acknowledgments}
		This research used facilities of the Brazilian Synchrotron Light Laboratory (LNLS), part of the Brazilian Center for Research in Energy and Materials (CNPEM), a private non-profit organization under the supervision of the Brazilian Ministry for Science, Technology, and Innovations (MCTI). The XDS beamline staff is acknowledged for their assistance during the experiments [Proposal 20180746]. We acknowledge Dr. R. Mendes Coutinho for fruitful discussions. This work was supported by Fapesp (Grants Nos. 2019/10401-9 and 2022/03539-7), CNPQ, ANPCyT (Grants PICT 2016/0204 and PICT 2019/02396), and CONICET (Grant PIP 2021/11220200101796CO).
	\end{acknowledgments}
	
	\bibliography{HoNiSi3}
	
\end{document}